\documentclass[12pt]{article}
\usepackage{graphicx}			
\usepackage{url}			

\newcommand{\yr}{$\phantom{1922}$}	

\begin{document}

\bigskip\bigskip

\centerline{\bf\large  ALLAN R.\ SANDAGE}
\medskip
\centerline{\bf 18 June 1926 -- 13 November 2010}
\medskip
\centerline{\bf Elected ForMemRS 2001}
\bigskip
\centerline{Donald Lynden-Bell CBE FRS}
\centerline{Institute of Astronomy, University of Cambridge}
\smallskip
\centerline{and}
\smallskip
\centerline{Fran\c cois Schweizer}
\centerline{Carnegie Observatories, 813 Santa Barbara St, Pasadena, CA}
\bigskip

\noindent
Allan Sandage was an observational astronomer who was happiest at a
telescope.  On Hubble's sudden death Allan Sandage inherited the
programmes using the world's largest optical telescope at Palomar to
determine the distances and number counts of galaxies.  Over many years
he greatly revised the distance scale and, on re-working Hubble's
analysis, discovered the error that had led Hubble to doubt the
interpretation of the galaxies' redshifts as an expansion of the universe.
Sandage showed that there was a consistent age of Creation for the stars,
the elements and the Cosmos.  Through work with Baade and Schwarzschild
he discovered the key to the interpretation of the colour--magnitude
diagrams of star clusters in terms of stellar evolution.  With others he
founded Galactic Archaeology, interpreting the motions and elemental
abundances of the oldest stars in terms of a model for the Galaxy's
formation.  He published several fine atlasses and catalogues of galaxies
and a definitive history of the Mount Wilson Observatory.
\let\thefootnote\relax\footnotetext{Accepted for publication in
{\em Biographical Memoirs of Fellows  of the Royal Society}\,.}

\bigskip\bigskip
\centerline{ANCESTRY  AND  EDUCATION}
\bigskip

\noindent
Allan Rex Sandage was proud to come from a mid-west farmer's family which
had moved from South Carolina to Indiana and, in 1881, to Lone Rock,
Harrison County, Missouri; he was born in Iowa City in 1926.
Allan was the only child of Charles H.\ Sandage, soon to become
a professor of business at Miami University in Oxford, Ohio, and Dorothy
Maud Sandage n\'ee Briggs, daughter of George Nathaniel Briggs, president
of Graceland College, Iowa (a Reformed Mormon foundation, now Graceland 
University).  Charles Sandage's father, Moses Sandage, was a farmer in
Lone Rock and had two other sons and a daughter, but Charles was the only
one who went to high school.  He proceeded to Graceland College and on to
the State University of Iowa, where he worked his way through graduate 
school to obtain a PhD in business administration.

Oxford was a small town in southern Ohio, and Allan grew up under austere
conditions during the Depression; orange juice was the height of luxury!
His was a university-oriented family with a father and mother who strongly
supported his will to learn.  They were not practicing Mormons, and on 
occasion he attended the local Methodist Church. His father had struggled
to get his education, so Allan always expected a hard road with four years 
of undergraduate work followed by four years to get himself a PhD in
science.  However, this was looked upon as a tremendous opportunity, a very
enjoyable part of life.  His interest in science was sparked when the family
moved  temporarily to Philadelphia where his father worked for the
government in 1936\,--\,37.  One of his boyhood friends there had a
telescope, and---on looking through it in the backyard---Allan was caught
by the wonder of astronomy.  Back in Ohio he read all he could find on
mathematics and science.  His schoolteachers realised his interest and
helped him both by after-class tuition and by suggesting further books.
``The Glass Giant of Palomar'' by David O.\ Woodbury was published in 1940
when Allan was 14.  This book contained a history of the Mount Wilson
Observatory, and he realised that it was possible to make astronomy one's
profession.  But at that stage he had no inkling that he would become the
most prominent user of the great Palomar reflector, whose construction was
delayed by World War II.  Eddington's books and Hubble's ``The Realm of the
Nebulae'' were other sources of inspiration.  In summer 1941 his father 
had a teaching appointment at Berkeley.  On the way there they took the 
opportunity of visiting both the optical shop at Santa Barbara Street and
the observatory itself on Mount Wilson.  

His father's war-work was in Boston, so his parents moved to Cambridge, 
Massachusetts.  But Allan, then 16, remained in Ohio with student 
accommodation at Miami University, where he majored in physics.  Professor
Ray Edwards headed a physics department of only three, but he instilled in
his students the importance of precision, dedication, excellence and
honesty  that remained hallmarks of Allan's work throughout his life. 

In 1944 he volunteered for the Navy and went into the electronics
maintenance programme for 18 months, first in Chicago, then at Gulfport,
Mississippi.  There he met many other scientists, including the astronomers 
Arthur Code and Albert G. Wilson, who was later responsible for the first 
Palomar sky survey with the 48 inch Schmidt. After a three-month training 
course, they were moved to Treasure Island in San Francisco  for a further 
nine months.

\begin{figure}
  \begin{center}
    \vskip -1.0truecm
    \includegraphics[scale=0.80]{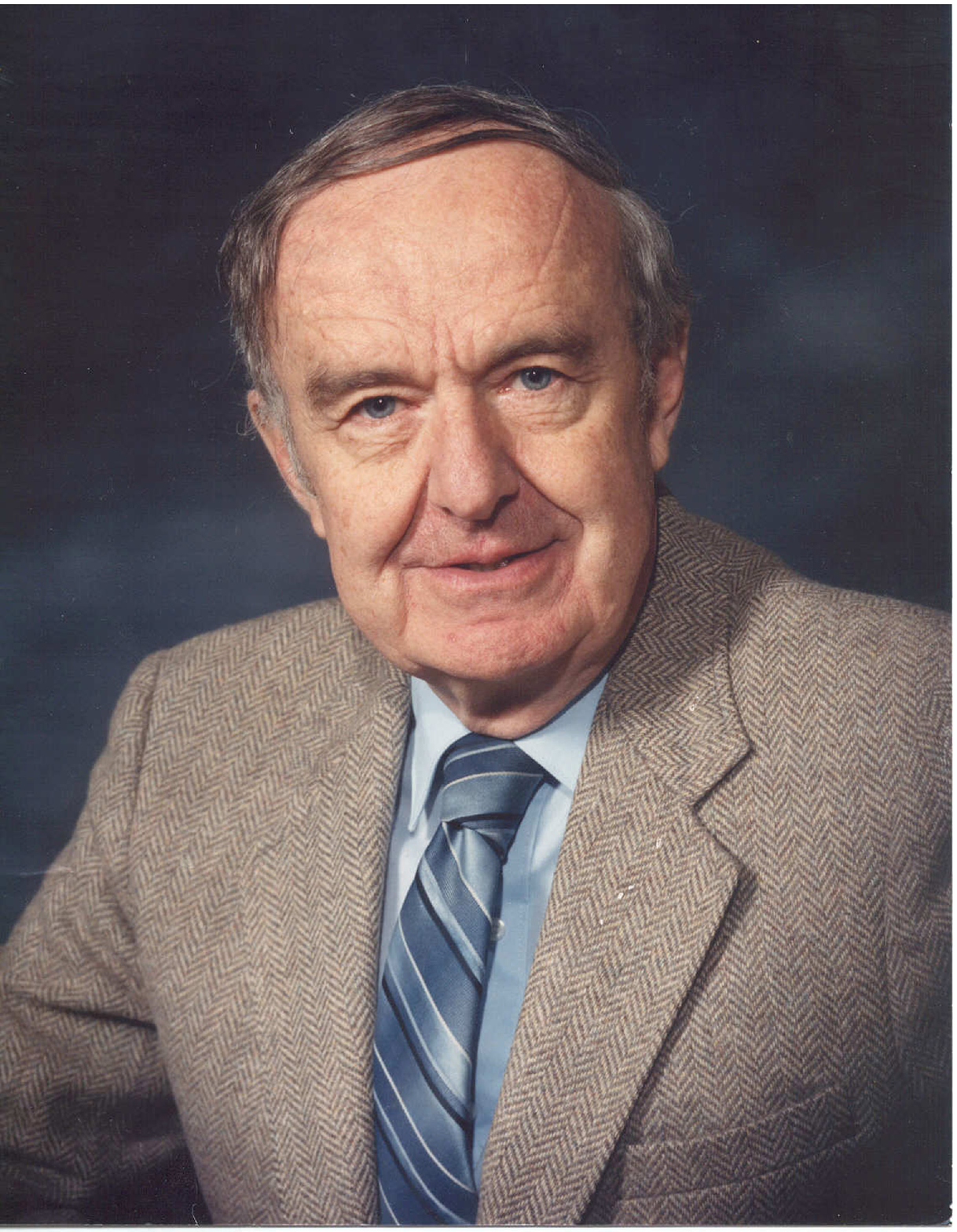}
  \end{center}
\end{figure}

At the end of the war his parents moved to Urbana, Illinois, where his 
father was appointed professor of advertising.  Living at home with the 
government benefits from the G.I.\ Bill made it possible for Allan to
resume his studies in the much larger physics department of the University
of Illinois.  In later years Allan had little contact with his father, who
became very prominent in advertising, but he remained very close to his
mother.  He became a heavy smoker, but when she died of cancer in 1970 he 
smoked his last cigarette.  Jeans's book ``Physics and Philosophy'' gave him
a strong interest in the philosophy of science, and this led him to further
reading in philosophy.  At Illinois Professor Robert Becker, who had been
at Caltech, taught analytical mechanics; at first Allan found the problems
too hard, but with Becker's help he mastered them.  There were no quantum
mechanics or relativity courses, but he got an excellent grounding in
electromagnetism, optics and mathematical physics.  He majored in 
physics and mathematics with a minor in philosophy, but meanwhile he started
a junior research problem in astronomy under Prof.\ Robert H. Baker.  The
techniques he learned there were very valuable for his later work at 
Caltech.  The project itself was part of a larger one co-ordinated by
Prof.\ Bart J. Bok of Harvard, who visited the observatory and invited
Allan to his observing summer school at the Agassiz station of Harvard.
There he met Vaini Bappu, who later did much for Indian astronomy, and
Halton (`Chip') Arp who will reappear.

In 1948 Allan applied to be a graduate student in physics at Caltech, but
in his letter of support Prof.\ Loomis, head of physics at Illinois,
mentioned his strong interest in Astronomy.  Lee DuBridge, replying to
Allan's application, said the graduate School in astronomy  was just
beginning and ``you have been accepted in that school at Caltech''. 
Although Fritz Zwicky was in Physics, Jesse Greenstein was the sole
professor of astronomy for those first years;
there were four graduate students and he taught all the courses: stellar
atmospheres, stellar interiors, the interstellar medium, practical
astronomy and astronomical methods of observation.  Allan was much in awe
of such erudition. 

Astronomy students were required to take physics courses.  Helmut Abt, a
fellow student who remained a staunch friend for life, recalls that he
and Allan puzzled hard over Prof.\ King's optics problems, at first
separately and then together, but got nowhere.  Later in class
Mr.\ Ferrell was called upon and immediately gave an elegant solution to
the first problem, then Mr.\ Parker likewise solved the second in two
lines.  Sandage and Abt felt outclassed.  It later transpired that Ferrell
was the only Caltech student with straight `A's all through, and Eugene
Parker became a world expert on magnetohydrodynamics at Chicago.  Sandage
claimed it was Abt's levelheadedness that helped him survive the courses.
Both had a hard work ethic that pulled them through.

Among the other students in their first year was Morton Roberts, who 
became a radio astronomer and, following Babcock's early work, produced  
good evidence for dark matter in the outer parts of the Andromeda Galaxy.  
It was this that drew Vera Rubin into that subject.  Among those in the 
following year was Chip Arp.  Harden McConnell, later a professor of 
chemistry at Caltech and then Stanford, writes of that time: ``We lived
in what was then called the greasy spoon, a somewhat tumbled-down frame 
building on campus that housed graduate students.  I bunked in the same
room as Donald Glaser, of bubble chamber fame; next door were Roy Craig,
a chemistry student, and Allan.  Graduate students of my vintage tended to
be rather sober.  The PhD qualifying exams were demanding and designed
to flunk a fraction of the class; research could be equally demanding.
Allan struck me as quite distinct from the other students.  I don't think
I ever saw him without a smile and something pleasant to say.  I never did
learn the source of his inner happiness, but it was certainly there."

\bigskip\bigskip
\centerline{THE  KEY  TO  STELLAR  EVOLUTION}
\bigskip

\noindent
Walter Baade was on the staff of Mt.\ Wilson.  During World War II the
darker sky due to the blackout in Los Angeles had enabled him to resolve 
the central parts of the Andromeda nebula and show that the brightest
stars there were red, like the brightest stars of the globular clusters.
This led to his concept of Population II.

Sandage and Arp asked Baade for a pre-thesis research project.  He
suggested locating the main sequence in the globular clusters.  Although
their brightest stars were far from the main sequence, it was assumed 
nevertheless that much fainter stars would probably lie on it.  Baade
taught the two graduate students how to use the 60 inch telescope on Mount
Wilson, and they took plates of the globular cluster M92.  Meanwhile, Baum
provided a faint photoelectric sequence in the cluster.  Arp and Sandage
shared the hard work of plate measurement, and within a year they had a
preliminary result.  Then Baade got Palomar 200 inch plates that went
deeper and that gave them a definite identification of the main sequence
and a rough estimate of its slope.

About that time Hubble asked Greenstein if there were a capable graduate
student who could help with his programme on galaxy counts.  Following a
suggestion of Eddington's, his aim was a better determination of $N(m)$,
the number of galaxies per unit area of sky of magnitude $m$, in order to
discover the geometry of space.  Hubble took plates on the 48 inch Schmidt
dithered so that each star gave a square of uniform blackening on the 
plate.  By measuring the blackenings on different plates, a photoelectric
sequence of stars of known brightness could be transferred from one area
to another.  Hubble wished to have reliable sequences down to the 18th
magnitude.  After getting Sandage started, Hubble departed for the summer
of 1950.  Sandage found inconsistencies of three tenths of a magnitude in 
the transfers and took on another short project under Greenstein while he
awaited Hubble's return.

Hubble did not return as expected.  He had suffered a serious heart attack
while fishing and needed time to recuperate.  When he came back he
wished to continue accumulating plates of the M81 group of galaxies to
identify Cepheid variables.  However, his health was too poor to endure
the long nights in the observing cage of the 200 inch telescope.  So, after
some initial instruction from Humason, the job of taking these plates fell
to Sandage, who later described his reaction:  ``Oh, it was fabulous.  I
can't really reconstruct the first three or four months of that period, so
much was happening.  First of all, it was an opportunity that was beyond
any imagination, observing with the 200 inch, and secondly, working on the
long-range programmes of cosmology with Hubble.  And at the same time 
being a graduate student, trying to pass the courses in physics and 
astronomy---so it was a very high-pressure atmosphere.  The work on the 
mountain was an escape, but you knew that your sins would catch up with
you, because you were four days away from campus and courses, and you were
pretty well swamped.  I'd bring the plates back and have a consultation
with Hubble.  He would then decide what to do next.''  Meanwhile, with
Baade as his adviser, he had decided that his thesis topic would be the
determination of the colour--magnitude diagram of the globular cluster M3.
With the work on M92 well on its way,  he already knew the methods, and M3
offered the prospect of seeing more of the main sequence.

Martin Schwarzschild (ForMemRS) was a great friend of Baade's and visited 
Mount Wilson every year.  The new results on M92 and the preliminary
results on M3 were very exciting to Schwarzschild, who had been working
theoretically on the evolution of the core of a star after it has burned
all its hydrogen.  He invited Sandage to come to Princeton and work on the
interpretation of his results for a year, once his observations were
complete.  Meanwhile, Sandage's work had so impressed Baade and Hubble
that, even before he completed his PhD, Bowen, the director, offered him a
staff position at Mount Wilson and Palomar Observatories.  This was a
dream come true, and in 1952 Allan set out for his post-doc at Princeton,
still without a PhD, but with his dream-job assured on his return.  The
eight months in Princeton turned his work on M92 and M3 into a seminal
discovery.  For the first time, the evolutionary tracks across the
colour--magnitude diagram fitted real data.  Sandage and Schwarzschild
understood that the turn-off point from the main sequence gave a method by
which stellar ages could be determined.  With their 1952 paper it became
clear that at last theories of stellar evolution had real power to explain
the facts.  (In 1963 Sandage and Schwarzschild  were jointly awarded the
Royal Astronomical Society's Eddington medal for this work.)

At the end of his time at Princeton, Sandage joined Baade, who took part 
in the summer school organised by Leo Goldberg in Ann Arbor Michigan.  
Later Sandage wrote a lively account of his return to California.  ``I
had been Baade's student during the previous four years, and we had come
to feel somewhat at ease with each other.  Baade had decided to buy a new
car at the factory in Detroit, to take delivery on the last day of the
school, and to then drive the long road back to California, asking me to
ride with him.  Ed Dennison drove us both to the factory pickup place for
a new 1953 Chevy.  After the car was in Baade's possession, Ed said he
would follow us for a few miles simply to see if all was OK.  We started
down the road out of Detroit toward California.  Baade (essentially a new
driver) was driving half on the road and half on the shoulder (sometimes),
or half on his side of the road and half on the other oncoming side,
alternatively.  Ed  signalled us to stop and enquired through the
rolled-down window if there was something wrong with the steering wheel.
Baade, sensitive, said all was OK, thank you very much, and we would be
on our way.  A few minutes later we unfortunately were!  Baade had no
sense of micro-compensation in steering; he would keep the wheel in a
rigid position until it was clear he had to change, not microscopically,
but grossly, long after it had become evident that it was required to do
so.  We went across the country for six days in large triangles, first
going toward the middle of the highway and then towards the corn or wheat
fields, or later towards the canyon drop-offs on the right.  The first
day was by far the worst because I was frightened.  I offered to drive,
but like all great men Baade believed in himself and thought only he could
save us from the oncoming drivers, who were astounded when we got close
enough to see their faces, and who Baade believed were simply poor drivers
that should be denied access to the road.  The trip became a bit easier
as it wore on because I could not help but sleep for most of the day,
avoiding the constant thrill of the road.  To save expenses, we had agreed
to share a double room in the motels along the way each night.  However,
Baade was a most accomplished snorer, so accomplished that it is simply
impossible to describe.  \dots\  After
two nights of no sleep for me, but restful sleep for the driver, I slept
away many of the driving hours.  But the real memory of that trip was the
conversation about astronomy on that cross-country adventure.  Baade, like
all scientists of substance, had a set view of how things were put
together, to be sure a view to be always challenged by the scientist
himself, but defended as well against all less informed mortals who
objected without simon-pure reasons.  The trip then became a riding
commentary on much of the world of astronomy and astronomers.  But that is
a rather different story than the magic of the summer of 1953, which in
many ways began the outside world's discussion of Baade's ideas that,
together with Schwarzschild's, spearheaded the modern understanding of
stellar evolution.''

Not long after Sandage returned to take up his new post at Mount Wilson 
and Palomar Observatories, Hubble died of a second heart attack.  Just as 
the mantle of Elijah fell on Elisha, bringing with it the awesome 
responsibility for Israel, so Hubble's cloak fell on Sandage, carrying
with it responsibility for furthering Hubble's cosmological programmes.
No other telescope could carry out such work, on which rested the
reputation of the world's greatest observatory.  Sandage knew he was a
good astronomer; now it was his duty to prove himself to be a great one.
With Schwarzschild he had already broken open the field of stellar
evolution.  Later, such understanding of the stars could be used to
further cosmology, but for the next decade his major research was devoted
to understanding stars. 	

\begin{figure}
  \begin{center}
    \vskip -0.2truecm
    \includegraphics[scale=1.4]{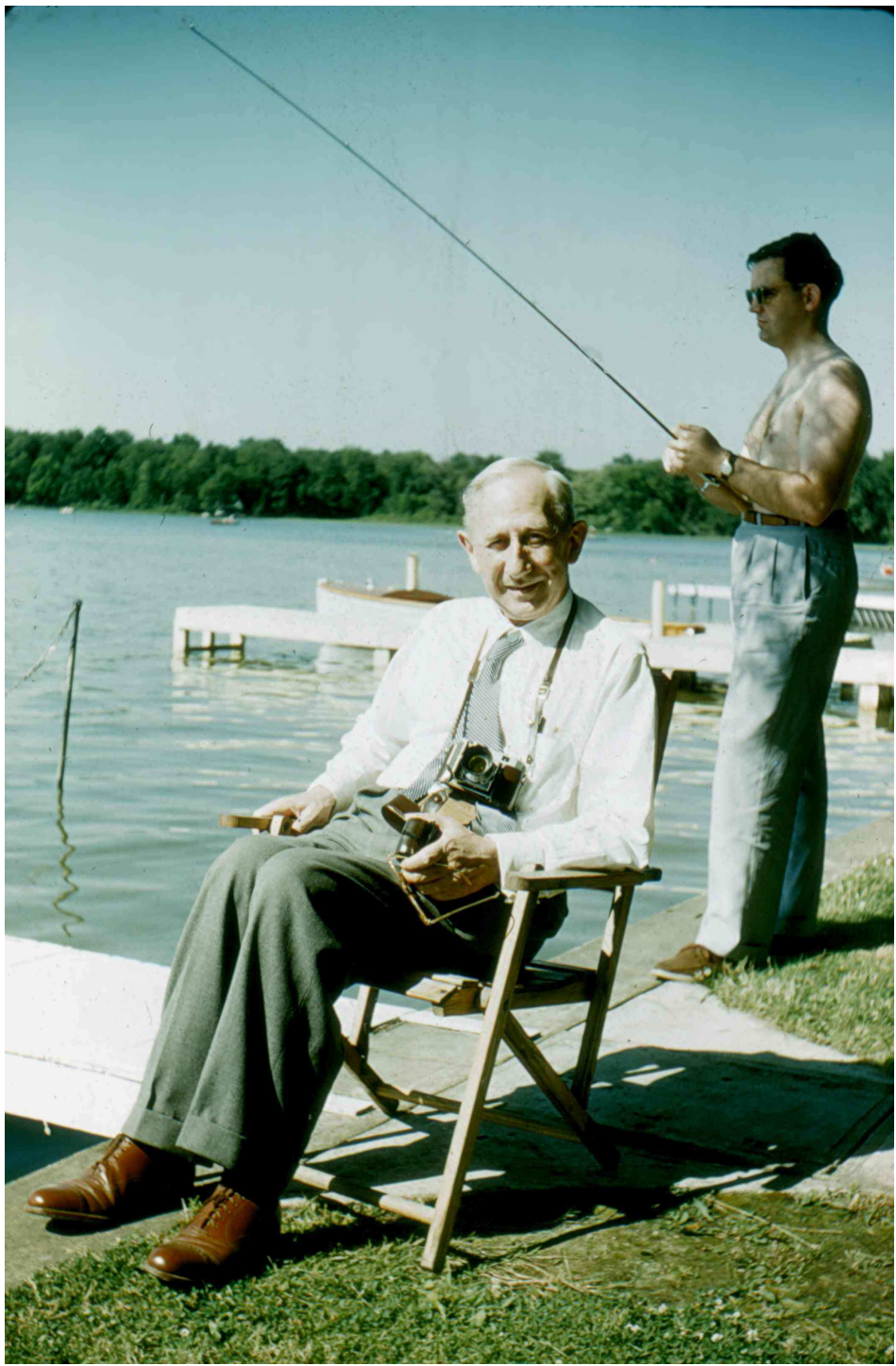}
    \vskip 0.3truecm
    \caption{
Walter Baade with Tom Matthews (fishing) at Bass Lake during the Michigan
summer school organised by Leo Goldberg in 1953.  Photograph by Owen
Gingerich.
    \label{fig1}}
  \end{center}
\end{figure}

At the Vatican Conference on Stellar Populations, organised by Father 
O'Connell, Sandage was the star performer, giving no less than four talks 
explaining the way that stars evolved and how their ages could be found 
from the colour--magnitude diagrams of their clusters.  Also in 1957, 
Sandage gave lectures at Harvard on the observational approach to stellar 
evolution.  In the audience was Mary Connelley, then a graduate student at 
Radcliffe, having taken her first degree at Indiana University in
Bloomington.  She had replaced Alice Farnsworth as head of astronomy at
Mt.\ Holyoke for three years, but she first met Allan much earlier in
1950, and again a year or two later at a conference in Michigan.  In 1959
Allan Sandage and Mary were married.  He often claimed that Mary had saved
his life.  When he came home from the office exhausted, disappointed,
furious or desperate, Mary would be kind, gentle and patient.  His temper
would soon cool. 

His close friends were Helmut Abt, Geoffrey Burbidge, Olin Eggen and 
Chip Arp.  Later, when Arp espoused the concept of large 
non-cosmological redshifts on flimsy evidence, that friendship broke for a
decade.  Allan felt responsible for getting Chip onto the Observatory
staff, and to be so unscientific was for him `sinning'.   Yet he never broke
with Geoffrey Burbidge who also espoused non-cosmological redshifts.

In 1958 Sandage visited South Africa with his friend Olin Eggen to work on
the Magellanic Clouds.  In Pretoria they were guests of the Thackerays.
Mary Thackeray had a strict rule against talking shop, which Sandage 
found quite a strain, but---though nervous---he enjoyed playing Father 
Christmas to the children.  At the invitation of the Astronomer Royal, Sir
Richard Woolley FRS, he also visited the Royal Observatory at Herstmonceux
Castle in Sussex, where later Eggen became chief assistant.

\bigskip\bigskip
\centerline{THE  COLLAPSE  OF  THE  GALAXY}
\bigskip

\noindent
In stars of normal metal abundance like the Sun, the atomic absorption 
lines crowd together at the ultraviolet end of the optical spectrum, 
blanketing out some of the light.  Metal deficient stars are relieved of
this blanket, so have an ultra-violet excess in comparison to normal
stars of the same colour.

Around 1959 Sandage and Eggen realised that this was a quick way of 
classifying stars by their metal abundances.  Following a lead found by
Nancy Roman, Eggen had been studying motions of nearby stars across the
sky and had produced a catalogue of stars with large transverse motions.
He and Sandage measured these stars photoelectrically and took spectra to
get radial velocities of those that lacked them.  Most stars in the solar
neighbourhood move in nearly circular orbits around the Galaxy and have
metal abundances not very different from the Sun, however the stars in 
Eggen's catalogue dived through the plane of the Milky Way and had orbits 
that took them both much further in and further out from the galactic
centre.  Some even went backward around the Galaxy.  They were all metal
deficient, and their colour--magnitude diagram showed that they were, 
like globular cluster stars, among the oldest in the Galaxy.

One of us (DL-B) was a post-doc working with Sandage at that time, and
we discussed what these findings meant.  Since the stars were among the
oldest, the rest of the Galaxy must have been gaseous then.  The gas could
not be so hot that it was supported against the Galaxy's gravity by
pressure, as it would then be far too hot to make stars.  Thus the gas
clouds must have been cool and followed the diving orbits still seen in
the stars.  But that is unsustainable; unlike the stars, the gas clouds
would collide.  In the 1962 paper widely known as ELS, we concluded that
the Galaxy formed in a dynamical collapse during which the metal-deficient
stars were formed.  But this collapse ended in starburst activity in which
the gas clouds collided, the rate of star formation rose to balance the
influx of gas and the metal abundance rose to values not far below solar.

ELS became the most cited single paper of each of its authors.  It laid
the foundations of what has come to be called galactic archaeology, which
now uses much more detailed abundances of the different elements and
correlates these with stellar motions.  More recent work on galaxy
formation puts emphasis on galaxies merging;  small galaxies have too
little gravity to hold in supernova debris, so cannot hang on to their
metals.  Such systems are still merging with the Galaxy, adding their
metal deficient stars to our halo.  The concept that the Galaxy initially
formed in dynamic collapse is nevertheless correct, because the emission
from a hot gas at galactic densities cools more rapidly than the
free-fall time.

\bigskip\bigskip
\centerline{QUASARS AND QUASI\,-\,STELLAR OBJECTS}
\bigskip

\noindent
One of the burning questions of the early 1960s was the nature of radio 
sources found by Ryle's group at Cambridge and by the Australians.  
Greenstein had convinced Caltech president Lee DuBridge of the importance
of radio astronomy, so money had been allocated to create the Owens 
Valley Radio Observatory (OVRO), and John Bolton, a Yorkshireman, had 
been brought over from Australia to get it going.  The Cambridge group had
made the observations for the third Cambridge catalogue.  The $2\times 5$
arc-minute positions were given to Henry Palmer of Jodrell Bank, who had
developed the radio-linked interferometer whose aerials were spread across
England and Wales.  This he used to determine the angular diameters of the 
radio sources.  Tom Matthews of OVRO attended a conference at which Palmer
gave his results, including about 30 sources with parts of such high
surface brightness that they were unresolved.  Thinking that these small 
sources might be the most distant and therefore the most interesting, 
Matthews asked Palmer for a list of those sources.  He returned to Caltech 
with this list, and Bolton readily agreed to a programme in which
R.\,B.\ Read refined the Cambridge positions at Owens Valley by using a
wider north-south separation of the dishes.  It was hoped that these
sources could then be identified optically.  Sandage took 200 inch plates
centred on the refined positions.  The possible identifications which
appeared to be galaxies were then given to Maarten Schmidt of Caltech, who
had taken over Minkowski's program on distant radio galaxies.

However, among the first identifications was 3C 48 in which a 16th
magnitude star lay in the error box.  Was this the first radio star?  In
October 1960 Sandage photometered the star and found it to be abnormally
bright in the ultraviolet.  He then took its spectrum.  It was quite
unlike anything he had seen before, just a number of emission lines
superposed on a continuous spectrum.  The wavelengths of the emission
lines made no sense.  They did not correspond to those of any known
element.  Sandage took his spectrum to Bowen, who had made his name by
identifying the nebular lines at 3726,\,3729\AA\ with forbidden lines of
oxygen.  Bowen could not interpret the spectrum, but said that Greenstein
had been studying the spectrum of highly ionised oxygen.  He was consulted
too, but still the spectrum made no sense. 

In the autumn of 1961 the Burbidges visited, and we all saw the direct 
plate of 3C 48.  Attached to the starlike image was an extremely faint
wisp, which could only be seen if the plate was viewed at a glancing
angle.  We all knew of the problem posed by the spectrum.  Quite soon
afterwards Sandage returned from observing and excitedly announced that
3C 48 must be a star.  It had varied by 0.3 magnitudes over a few months,
far too rapidly for a galaxy of several thousand light-years in diameter.
This larger variation convinced him that the small night to night
variations he had seen in January 1961 were definitely real. Other
mysterious radio sources soon followed.  3C 286 and 3C 147 had fewer
emission lines, but also had ultraviolet excesses.  Eventually a lunar
occultation observed in Australia gave a very accurate position for
3C 273, and the position coincided with a 13th magnitude star.  Schmidt
took the spectrum, which had numerous unfamiliar emission lines.  On being
pressured to publish by the radio astronomers, Schmidt finally saw that,
if he ignored the brightest lines, the others were spaced not unlike the
Balmer series.  The Balmer series they were indeed, but with a redshift
of 0.158.  Never before had such a bright object been found with such a
large redshift.  Greenstein then understood the spectrum of 3C 48 and
published its redshift with Matthews.  To Greenstein the discovery was a
vindication of his founding of Owens Valley and an emergence of his
Caltech astronomy department from being overshadowed by the expertise of
Mount Wilson, but Sandage, who regarded 3C 48 as his object, was omitted
from the party.  He had been left uninvited by the very professor whom
he so admired as his teacher.  However, he continued to study further
quasars, and the next major discovery was his. 

What Sandage discovered was that the quasars were not isolated rarities, 
but the radio-loud members of a much larger set of objects (QSO) which 
had relatively little radio emission.  We know of no other paper that was 
received by the Journal on the day of its publication, but Subramanian 
Chandrasekhar (FRS), who was editor of the Astrophysical Journal, had a
high opinion of Sandage and held up publication so that his paper 
announcing the quasi-stellar galaxies could be included.  This paper was
written in the excitement of discovery.  Sandage failed to allow for a
significant contamination of the number counts at the brighter magnitudes
by white dwarfs.  This caused controversy until their contribution was
sorted out via spectroscopy.  However, Sandage's basic point was correct;
QSOs far outnumber quasars.  In 1969 his result was used by DL-B to
estimate the number of dead QSOs.  He concluded that they were probably
giant Black Holes and roughly as numerous as large galaxies.  On hearing
his theory that galactic nuclei were collapsed old quasars, Sandage wrote
to get him an extra invitation to the 1970 Vatican conference on the
Nuclei of Galaxies.

More recent work shows QSOs to be strong in X-rays and even in gamma rays,
and it is now accepted that they are powered by material that is heated
as it spirals down into black holes of many million solar masses.  In this
sense, Sandage was the discoverer of the large number of giant Black Holes
in the Universe.

\bigskip\bigskip
\centerline{THE AGE OF CREATION}
\bigskip

\noindent
When Sandage and Schwarzschild first realised that  main-sequence 
turn-off points allowed them to find the ages of star clusters, the age of
the Earth was thought to be about three billion years.  They hoped to get
a similar figure for the old clusters.  Baade's work on M31 had shown it
to be at about twice Hubble's original distance estimate, but Hubble's
estimates of other distances and redshifts gave a timescale of only
1.8 billion years.  Soon after his return from Princeton, Sandage
inspected the best 200 inch plates.  He found that many of the objects in
other galaxies that Hubble had thought were their brightest stars, were
now resolved into tight knots of several bright stars embedded in ionised
hydrogen regions.  Hubble, by assuming that these were single stars
similar to the brightest stars of the Milky Way, had attributed to them
too low a luminosity and had therefore got his distances too small.  Thus
Hubble's discrepancy between the time at which the galaxies were close
together and the age of the Earth was now removed, and the age of the old
star clusters seemed to agree also.  Sandage was delighted by this
consistency of three different methods of age dating, which was in tune
with his strong belief that science is a consistent body of knowledge. 

\begin{figure}
  \begin{center}
    \includegraphics[scale=0.75]{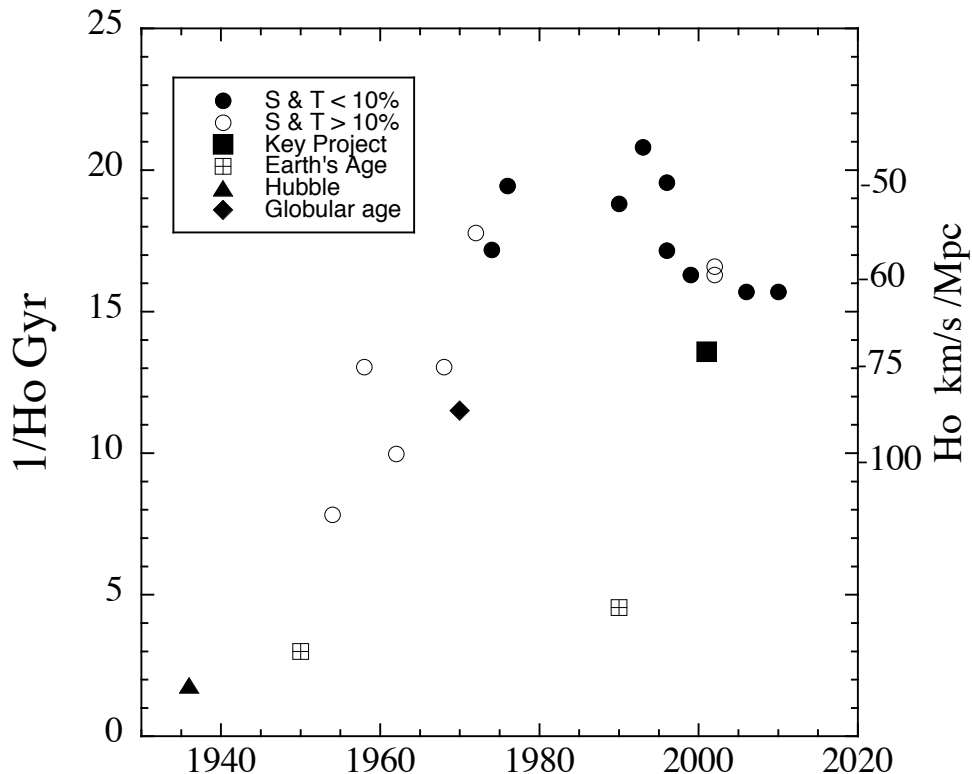}
    \caption{
Sandage and Tammann's (S \& T) estimates of the inverse Hubble constant,
$1/H_0$, plotted versus the year of the estimate.  Four corresponding
values of $H_0$ itself are marked on the right.
    \label{fig2}}
  \end{center}
\end{figure}

However none of the methods was secure.  The estimated age of the Earth 
is now 4.6 billion years.  Hubble's pioneering estimates of galaxy
distances could be improved with photoelectric magnitudes of faint stars
and by finding cepheids in other galaxies, and the age of the globular
clusters from stellar evolution depended on the poorly-known masses of
the stars at the turnoff point.  Furthermore, both the helium and the
metal abundances in those stars affect the age determination from stellar
evolution theory.  Perhaps it is not too surprising that these ages have
over the years ranged up to 18 billion years, although no modern estimate
is less than 10 billion.  Sandage derived 11.5 billion years for his
beloved M3 and two other globular clusters.  In Figure 2 we have plotted
different  estimates of the inverse of the Hubble constant, which is a
time, against the date at which the estimates were published.

Most stars are too dim to be seen in other galaxies.  Thus only the very 
rare, extremely bright stars can be used as ``standard candles''.  The 
problems of measuring distances to them are:  Firstly, can we recognise 
precisely the same type of stars, and how accurately do these stars have
the same luminosity?  Secondly, can we find similar stars in our galaxy
and measure the distances accurately through the murk of dust in which we
find ourselves?  Cepheid variables are very valuable, as their periods can
be measured accurately.  Cepheids of a given period have a quite
well-defined luminosity.  The dimmer short-period Cepheids are common
enough that the distances to some in our galaxy can be measured directly.
The Pole Star is one such, but it is only the brighter ones that are
useful to measure the large distances needed for determining Hubble's
constant.  There are a few bright Cepheids in the Milky Way whose
distances can be measured with the desired accuracy.  For many years these
problems were circumvented by using the Large Magellanic Cloud (LMC), our
nearest satellite galaxy, as an intermediary.  It has both the bright
long-period Cepheids and the dimmer short-period ones, so these can be
calibrated against one another.  Since the 1980s, with the advent of CCDs,
it has been possible to measure LMC stars as faint as the main sequence.
However in the Galaxy and in other large galaxies the stars have greater
metal-abundances than in the LMC, and the luminosities of Cepheids of a
given period depend on the metal abundance.  Sandage and his main
collaborator, Gustav Tammann of Basel, wrestled with such problems for
45 years.  Gradually, instrumental improvements, the coming of CCDs rather
than photographic plates, and the high resolution of the Hubble Space
Telescope (HST), which allowed more Cepheids to be identified and
measured, eased the problems. 

The measurement of Hubble's constant had been singled out as a key project
for the HST, but the project was shared over a large team which Sandage
and Tammann did not join.  The team effort eventually found a Hubble 
constant of $72\pm 3$ (random) $\pm$ 7 (systematic) km/s/Mpc.  In their
last paper, Sandage and Tammann did not agree, finding  $62.3\pm 1.3$
(random) $\pm$ 5.0 (systematic) km/s/Mpc.  Time will tell which estimate
is closer to the truth.  The work has been highly productive also in
isolating good secondary distance indicators, which could be used to great
distances.  First Sandage, often supported by Jerome (`Jerry') Kristian,
found that the brightest members of clusters of galaxies were quite good
standard candles, but the real gem came when Sandage and Tammann worked
to turn supernovae into  standard candles, an idea that Sandage attributed
to Zwicky.

\bigskip\bigskip
\centerline{TYPE Ia SUPERNOVAE}
\bigskip

\noindent
A 1968 paper by Kowal, who worked with Zwicky and Wallace Sargent (FRS) to
discover supernovae, found that supernovae of type I reached the same 
luminosity at maximum to within 0.61 magnitudes.

Sandage and Tammann realised this could be further refined by restricting
consideration to supernovae of type Ia.  Such a bright standard candle
could eliminate many of the intermediate steps in the determination of the
distance scale.  In 1991 Sandage therefore wrote an observing proposal 
for the Hubble Space Telescope to calibrate via Cepheids the distances to 
those nearby galaxies in which supernovae of type Ia have been observed. 
The brightest classical novae decline from maximum light faster than the
dimmer ones, but for supernovae Mark Phillips discovered that the dimmer
ones decline faster.  This correlation allowed the luminosity at maximum
of a type Ia to be predicted to within 0.2 magnitudes, and this spread can
be even further reduced by using the colour.  To carry out the detailed 
work of Cepheid discovery and measurement Sandage and Tammann asked 
others to join them, and Abi Saha was delighted to join.  He was the lead 
author on the HST Cepheid work, which lasted fifteen years and found many 
Cepheids in each of the dozen galaxies studied.  In their final paper
summarising the results of this campaign, Sandage, Tammann and Saha---using
the best local Cepheid calibration available---derived a Hubble constant of
$62.3\pm 5.2$ km/s/Mpc.  They point out that the difference between this
value and that of the Key project is primarily due to their use of this
different local calibration.

\bigskip\bigskip
\centerline{THE SOUTHERN OBSERVATORY}
\bigskip

\noindent
In 1964 Horace Babcock and Sandage went to Bowen, the director, and
proposed that the future of the observatory lay in expansion into the
Southern hemisphere.  This had been a dream of Hale's from the early days.
Bowen felt there was too much to be done to ensure the success of the
Palomar 200 inch by improving its instrumentation and that of the
Mt.\ Wilson telescopes.  Any Southern Observatory would divert effort from
that.  However, Babcock and Sandage pressed their point, and eventually
Bowen agreed to fund site-testing in the South.  This was organised by
Babcock.  It soon became apparent that the European Southern Observatory's
choice of Chile was a very good one, but the site-testing still took five
years.  Eventually the Carnegie team chose Las Campanas, a site somewhat
higher and more remote than ESO's La Silla.  Babcock and Sandage made the
case for a southern 200 inch telescope in collaboration with AURA (the
Association of Universities for Research in Astronomy).  This got very 
close to being funded by a large grant from the Ford Foundation together
with the NSF, but, just as it was being agreed upon, a new head of the
Ford Foundation was appointed who had other views.  Eventually both large
grants went to AURA without Carnegie involvement. 

Sandage flew back from his sabbatical in Australia and joined Babcock to 
meet Crawford Greenewalt, a trustee of the Carnegie Institution and
chairman of its astronomy subcommittee.  Greenewalt, being a scientist 
and engineer, understood the importance of the Chile site.  From this
meeting came a strong commitment by Greenewalt to support the
Carnegie Southern Observatory (CARSO), if a suitable site was found.
Although many of his staff thought Chile was rather remote and were not
in favour of diverting funds there, Babcock was rightly convinced that
this was the frontier where the observatory's future lay.  In late
1968 he met President Eduardo Frei and, to his surprise, obtained instant
permission to purchase nearly 100 square miles of land including Cerro
Las Campanas.  The Carnegie Institution purchased the tract within months.
Henrietta Swope had already given the bulk of her inheritance anonymously
to develop the site and put a 40 inch telescope there, which was
inaugurated in 1971 and bears her name.  In lieu of an unaffordable 200
inch, Babcock and engineer Bruce Rule advocated a lesser telescope, and
a generous gift by Crawford and Margaretta Greenewalt made possible the
completion in 1976 of the 100 inch Ir\'en\'e du Pont telescope, named in
honor of Mrs.\ Greenewalt's father.  Thus by the time Babcock left as
director, the du Pont was established on one of the world's best sites.
He, Bowen, and Arthur Vaughan had insisted on very good optics to match
the excellence of the seeing.  When it was first commissioned, there were
no spectrographs yet, but a remarkably wide field covering 2.2 square
degrees at the Cassegrain focus.  With many of the staff less than
enthusiastic about going all the way to Chile and many awaiting the
spectrographs, Sandage felt he must show what the telescope could do in
its present form.  He first completed his programme for the Revised
Shapley Ames catalogue and then devised the Virgo cluster survey,
described below.

Sandage greatly admired Babcock's foresight and tenacity in getting Las 
Campanas built.  However this was achieved when funds were scarce, and the
Caltech astronomers felt that Palomar needed more instruments, those on 
Mount Wilson were old and there was too much light pollution from Los 
Angeles.  Unsurprisingly, they had the feeling that they were sharing
their magnificent 200 inch telescope with the Carnegie astronomers and
were not getting much back.  There was also friction over appointments.
Because the telescopes were operated in common, both staffs had a say in 
staff appointments and they sometimes disagreed.  By the time Maarten
Schmidt succeeded Babcock as director in 1977, tensions between the
Carnegie and Caltech staffs made the joint observatory meetings
contentious.  Schmidt decided that the problems would be best resolved by
dissolution of the union that had run the observatories jointly for
30 years.  Sandage opposed the breakup, which he considered unnecessary,
but afterwards he broke off relations with Caltech.  It is interesting to
speculate that---had the union lasted a few years longer, so that more
Caltech staff experienced the wonderful conditions in Chile---then perhaps
the differences could have been resolved with freedom to appoint staff,
but no formal split.

The growth of Las Campanas meant that Carnegie funds remained scarce, and
despite the Mt.\ Wilson site having better seeing than Palomar, the
brighter sky led to less pressure to update the equipment there.  Finally,
to Sandage's horror, Mount Wilson---the home site at which the history of
20th-century astronomy was concentrated---was abandoned by the Carnegie
Observatories.  Sandage, with his collaborator Gary Fouts, used it to the
last, securing a much better database than that available to ELS.  They
confirmed a significant thick-disc population of stars with between one
tenth and one third the metal abundance of the Sun, which had been
advocated by Gilmore and Reid.  Sandage's very strong feeling for the
history of astronomy meant that he was particularly hurt by the decision
to abandon Mount Wilson.

\bigskip\bigskip
\centerline{GALAXY CLASSIFICATION AND SURVEYS}
\bigskip

\noindent
Upon Hubble's death in 1953, Sandage inherited his unique collection of
photographic plates of nearby galaxies obtained with the large Mt.\ Wilson
reflectors.  With it fell upon his shoulders the task of carrying out
Hubble's intended revision of his initial galaxy classification system of
1926.  This initial classification---based on earlier efforts by Max Wolf
in Heidelberg and Heber Curtis at Lick Observatory---had proposed three
main classes of extragalactic `nebulae': amorphous ellipticals, spirals
and irregulars, and had found its culmination in the famous `tuning-fork
diagram' published in ``The Realm of the Nebulae'' (1936).
Guided by notes and fragments of a manuscript written by Hubble, plus
memories of discussions with him, Sandage undertook the revision.
He supplemented the Mt.\ Wilson plate collection with many new plates
obtained at the Hale 200 inch and Schmidt 48 inch telescopes on Palomar.
After extensive visual inspection of plates old and new, he described the
revised morphological classification system in ``The Hubble Atlas of
Galaxies'', published in 1961 by the Carnegie Institution of Washington
and lavishly illustrated with photographs of 176 individual galaxies.
His detailed description of the revised Hubble types included crucial new
evidence for the existence of Hubble's S0 galaxies, a hypothesized 
transition type between the amorphous, red and dead elliptical galaxies
and the bluish disk galaxies with their splendid spiral patterns.
The Hubble Atlas remains one of Sandage's most cited publications to date;
coming at the right moment after the dissemination of the Palomar
Observatory Sky Survey prints, it made large-scale photographs of galaxies
obtained with the world's largest telescopes available to all astronomers,
enabling them to type galaxies seen on the prints by comparison with the
prototypes of the revised Hubble classification.
	
In 1975, Sandage supplemented his classification with an extensive review
of various galaxy classification schemes developed by astronomers
throughout the 20th century.  Though rarely cited, his review chapter in
Volume IX of {\it Stars and Stellar Systems} is still well worth reading.
The landmark volume  itself, entitled ``Galaxies and the Universe'', was
co-edited by Allan with Mary Sandage and Jerry Kristian.  Sandage was well
aware of the great potential value of uniformly conducted surveys.  He
admired the Harvard all-sky survey of 1246 bright galaxies, down to 13th
photographic magnitude, published by Shapley and Ames in 1932, and decided
early on to collect and measure redshifts for all these galaxies and to
reclassify them on the revised Hubble system.  A crucial part of this 
effort was the imaging of all Shapley-Ames galaxies south of declination
$-15$ degrees, the general limit of Hubble's Mt.\ Wilson galaxy plate
collection.  Beginning in 1974 with the Swope 40 inch telescope at Las
Campanas, and continuing with the du Pont 100 inch telescope after its
1977 commissioning, Sandage photographed as many of the southern galaxies
as possible, classified them, and in 1981 published---with Gustav
Tammann---the ``Revised Shapley-Ames Catalog of Bright Galaxies'', often
referred to as RSA.  Besides a complete listing of all Shapley-Ames
galaxies giving their revised Hubble types and measured redshifts, the RSA
contained high-quality photographs of 84 type-defining galaxies arranged
in 15 panels.  A novel feature of the catalog was that it included
intrinsic-luminosity classes, on a system first proposed by van den Bergh,
for most spiral galaxies.  In 1987 Sandage and Tammann published a second
edition of the RSA with improved types for about 200 galaxies, based on 
new Las Campanas plates.  If the two editions of the RSA are combined,
they replace ELS as Sandage's most cited publication.  To accompany and
supplement the RSA, Sandage and Bedke published in 1994 ``The Carnegie
Atlas of Galaxies'', two massive volumes showing high-quality photographs
of 1168 Shapley-Ames galaxies obtained with the large reflectors at Mt.\
Wilson, Palomar, and Las Campanas.  In his preface, Sandage commented:
``Given the evident responsibility to preserve this unique photographic
record, we deemed that the way to make the collection available for widest
possible use was to compile this atlas.''
 
A surveyor of galaxies at heart, Sandage also exploited the exceptionally
wide field of view ($1.5\times 1.5$ degrees) and excellent plate-scale
(10.8 arcsec/mm) of the du Pont telescope to conduct a survey of the Virgo
cluster of galaxies from Las Campanas.  The survey consisted of 67 large
($20\times 20$ inch), blue-sensitive photographic plates covering an area
of about 140 square degrees centered on the famous nearby cluster.  It
yielded a catalog of 2096 galaxies, of which 88\% were judged to be likely
or possible cluster members.  In six papers published 1984\,--\,87 Sandage 
and his collaborators (mostly Tammann and Bruno Binggeli) described the
properties of Virgo cluster galaxies, introduced a new classification
system for the dwarf members, and determined the structure and kinematics
of the cluster itself.  A key conclusion, congruent with developing
knowledge of galaxy clusters, was that the cluster core and envelope are 
both still forming, whence the Virgo cluster is young.

\bigskip\bigskip
\centerline{THE REALITY AND UNIFORMITY OF THE HUBBLE EXPANSION}
\bigskip

\noindent
Hubble to his dying day was very reticent about the expansion
interpretation of the redshifts of galaxies.  He always claimed that the
numbers of galaxies at a given magnitude would not show a uniform universe
if he adopted the expansion interpretation.  It was Sandage's reworking of
Hubble's data after his death that resolved this difficulty.  Hubble had
omitted a factor of $(1+z)$ in correcting his magnitudes, and it was this
that led to his contradiction.  The removal of that contradiction did not
of itself demonstrate the reality of the velocity interpretation of the
redshift.  One test is readily demonstrated.  The redshift is independent
of wavelength.  Tolman devised another direct test, and Sandage undertook
this test first with Jean-Marc Perelmuter and then with Lori Lubin.
Surface brightnesses should appear reduced by the factor $(1+z)^{-4}$ when
they are observed at redshift $z$. Lubin and Sandage carried out this test
and, assuming the form $(1+z)^{-a}$, they found $a = 4\pm 0.4$ in good
agreement with Tolman's prediction.  This result was only obtained after 
a somewhat uncertain allowance had been made for the gradual dimming of
elliptical galaxies as they evolve, but even without that correction the
``tired-light'' theory, which predicts $a=1$, does not fit.

After Hubble's death, Sandage joined Humason and Mayall from Lick 
Observatory in a great paper that described the state of Cosmology in
1956.  Sandage wrote the theoretical part of this paper.  By then
over 800 extragalactic redshifts were known as a result of long
hours at the telescope.  Forty years earlier Vesto Slipher at the Lowell
Observatory at Flagstaff obtained some 30 redshifts, typically exposing
for three nights for each one!

In his early papers on cosmology, and in particular his 1961 paper ``On 
the Ability of the 200 inch Telescope to Discriminate Between Selected 
World Models'', Sandage laid emphasis on the attempt to find the curvature 
of the redshift--magnitude relation as a way of measuring the expected 
deceleration of the expansion due to gravity.  Several times he reported 
marginal detections of such curvature in the expected direction, but all
of these proved ephemeral.  Later he realised that the density of the
universe can be measured more easily from deviations from the smooth
Hubble flow due to the significant gravity of large-scale concentrations
of galaxies, such as the Virgo cluster.  Sandage had always expected the
density of matter to be close to the density at which its gravity causes
the Universe to close.  He was constantly surprised by the smoothness of
the Hubble flow, so that even large concentrations of matter produced
considerably smaller changes in the velocity field than he expected.
Somewhat reluctantly he concluded that there was too little matter to
close the Universe.  Here he was quite correct.  However, his intuition
that the Universe should be on the edge of closure was also correct.  It
is dark energy, not matter, that produces most of the closure density. 

When at last there were sufficient supernovae of type Ia with redshifts of 
about a half, other groups did find the curvature in the redshift--magnitude 
relation, but in the opposite sense, corresponding to a universe whose 
expansion is accelerating.  This surprising result was soon agreed on  by
the two groups studying high-$z$ supernovae, but it was also strongly
backed by results from the Boomerang balloon experiment around Antarctica.
These showed the first peak in the spectrum of fluctuations over the sky
of the Cosmic Microwave Background, which clearly gave a value close to
the closure density.

\newpage
\centerline{THE CENTENNIAL HISTORY OF MOUNT WILSON OBSERVATORY} 
\bigskip

\noindent
From his early reading of ``The Glass Giant of Palomar'' Sandage had found 
inspiration from the history of Hale's Observatory on Mount Wilson.  He
met and knew well some of the first generation astronomers, Hubble and
Humason in particular, and shared stories with others at the monastery. 
Many of the second generation were still on the staff when he joined in 
1953.  He had a deep feeling for the magic mountain, where he had started
his communing with the stars, where Michelson had measured the
velocity of light and, with Pease, had developed his stellar
interferometer.  This was the place where Hale's laws on the magnetism of
sunspots were discovered and Babcock at last measured the Sun's magnetic
field and explained its cycle, where St John had looked for Einstein's
gravitational redshift in the solar spectrum and where Adams had found it
in the spectrum of the white dwarf Sirius~B.

When the Carnegie Institution decided to publish a history of its work
during its first century, Sandage was clearly the person to ask to write
the volume on the history of the observatory.  He responded with enthusiasm
and produced a magnificent scholarly work, in which the whole development
of 20th-century astronomy is intertwined with the history of the
observatory which led the subject.  Only he had the detailed knowledge of
the staff and their work.  He had heard their anecdotes while eating
midnight lunches on the mountain.  In this work he shows his deep love for
the institution in which he spent all his working life, his belief that
many of the early contributions from Mt.\ Wilson have not been fully
recognised and his deep anguish at Carnegie's withdrawal from Mount Wilson
in 1985.  This history ends in the 1950s.  Initially he had planned to put 
the later history into a second volume, but the effort of completing the 
650 pages of the first sapped his energy, and he left only fragmentary
notes for any sequel.  He had already written articles on the first fifty
years of Palomar.  Hale's farsightedness shines through the volume.  With 
the new developments planned for Las Campanas, the future of his
observatory is bright and well in keeping with his dictum ``Make no small
plans''.

\bigskip\bigskip
\centerline{THE  MAN}
\bigskip

\noindent
Allan Sandage was a tall powerful courteous man with a ready smile and 
strong blue eyes.  He had a natural presence and a slow incisive delivery. 
This, coupled with his encyclopaedic knowledge of astronomy and its
history, made him a dominant participant in those conferences that he
attended.  These became fewer over the years; he preferred to work and
find out things, rather than debate their reality.  His attitude was that
debates settle nothing except who is the best debater, and the true
answer is always decisive evidence gained at the telescope.

Allan and Mary had two sons, David and John, and later three grandchildren.
In most of the early years the family took camping holidays.  They were
all musical, and in the long nights in the 200 inch observing cage Allan
liked to listen to opera, especially Wagner.  The whole family particularly
enjoyed the sabbatical year they spent on Mount Stromlo in Australia in
1968--69, when the boys travelled down to school in Canberra.  They even
extended their stay to 15 months.  More recently they went on cruises,
often down the Mississippi; these were sometimes timed so that Allan could
avoid a conference or a meeting that he did not wish to attend.  He had a
strong conviction that the world with its natural laws was not created by
chance and must be there for some purpose.  He explored religion, but did
not find it satisfied his quest for knowledge.

In his later years he withdrew himself into his work, in order to avoid
debate and controversy.  While this resulted in a phenomenal output of
papers, to some degree it deprived the younger generation of knowing the
fun, the charm and the historical perspective of a great astronomer.
Diagnosed with pancreatic cancer in the Fall of 2009, Allan Sandage still
managed to take his family on one last cruise to Hawaii, do research at
home, and publish three first-authored papers during his last twelve
months.  He died on 13 November 2010, surrounded by his loving family,
at his home in San Gabriel, California.

\bigskip\bigskip
\centerline{HONOURS  AND  AWARDS}
\bigskip

\noindent
In the inaugural year of the Gruber Prize for Cosmology two full prizes
were awarded; one to Sandage was for Observational Cosmology.
He received the Crafoord Prize of the Swedish Academy of Sciences in
1991, and the President's National Medal of Science in 1971.
Other medals were the Eddington (1963) and Gold (1967) medals of the
Royal Astronomical Society, the Pope Pius XI gold medal of the
Pontifical Academy of Science in 1966, the Franklin Institute's Elliott
Cresson medal (1973), the Bruce medal of the Astronomical Society of the
Pacific (1975), and the Adon Medal of the Nice Observatory (1988).
He was Russell Lecturer to the American Astronomical Society in 1973 and 
received their Helen Warner prize in 1960.  He received the Tomalla
Gravity prize of the Swiss Physical Society in 1993.

Sandage was elected to the US National Academy of Science in 1963, but
resigned in 1980 when the Academy failed to elect his friend and
collaborator Olin Eggen.  He was a Member of the Lincei National Academy
in Rome.  He was elected a Foreign Member of the Royal Society in 2001.
He had honorary degrees from at least eight different universities and
colleges.

Finally, main belt asteroid 9963 Sandage, originally 1992 AN and
discovered by Eleanor Helin, is named in his honour.

\bigskip\bigskip
\centerline{ACKNOWLEDGEMENTS}
\bigskip

\noindent
{\small
Many sent us reminiscences of Allan Sandage; several we have used
explicitly, the spirit of others we have tried to incorporate into the
text.  In particular we thank Mary Sandage for answering our questions and
giving us details that only she knew.  Allan's greatest collaborator,
Gustav Tammann, sent much, as did Owen Gingerich, Helmut Abt and John
Kormendy.  We thank Harden McConnell, Michael Feast, Tom Kinman, Ken
Freeman, Vera Rubin, Rob Kennicutt, Harry Fergusson and Harry Nussbaumer
for material that helped us.  Sandage's account of his road-trip with
Baade was published in an article by Owen Gingerich in {\it Physics Today}
{\bf 47}, no.\ 12, pp.\ 34--40, 1994.  The paragraph describing his
reaction to the request that he take 200-inch plates for Hubble was taken
from an oral history interview by Dr.\ Spencer Weart on May 22, 1978, at
Santa Barbara St., Pasadena, California, which we found particularly
useful in preparing this memoir.  See interview with Dr.\ Allan Sandage,
Niels Bohr Library \& Archives, American Institute of Physics, College
Park, Maryland, USA:\ \
\url{http://www.aip.org/history/ohilist/4380_1.html} }


\bigskip\bigskip
\centerline{SELECTED BIBLIOGRAPHY}
\medskip

\begin{description}
\itemsep -2pt
\small
  \item[1952] (With M.\ Schwarzschild) Inhomogeneous stellar models. II.
		Models with exhausted cores in gravitational contraction.
		{\it Astrophys.\ J.} {\bf 116,} 463--476.
  \item[1953] (With H.\ C.\ Arp \& W.\ A.\ Baum) The HR diagrams for the
		globular clusters M92 and M3.
		{\it Astronom.\ J.} {\bf 57,} 4--5.
  \item[\yr]  The color--magnitude diagram for the globular cluster M3.
		{\it Astronom.\ J.} {\bf 58,} 61--75.
  \item[1956] (With M.\ L.\ Humason \& N.\ U.\ Mayall) Redshifts and
		magnitudes of extragalactic nebulae.
		{\it Astronom.\ J.} {\bf 61,} 97--162.
  \item[1958] In {\it Stellar Populations}, Vatican Symposium, ed.\
		D.\ J.\ K.\ O'Connell.
  \item[1961] The ability of the 200-inch telescope to discriminate
		between selected world models.
		{\it Astrophys.\ J.} {\bf 133,} 355--392.
  \item[1962] (With O.\ J.\ Eggen \& D.\ Lynden-Bell) Evidence from the
		motions of old stars that the Galaxy collapsed.
		{\it Astrophys.\ J.} {\bf 136,} 748--766 (ELS).
  \item[1963] (With T.\ A.\ Matthews)  Optical identification of 3C 48,
		3C 196, and 3C 286 with stellar objects.
		{\it Astrophys.\ J.} {\bf 138,} 30--56.
  \item[1965] The Existence of a major new constituent of the universe:
		The quasi-stellar galaxies.
		{\it Astrophysical J.} {\bf 141,} 1560--1578.
  \item[1978] (With J.\ Kristian \& J.\ A.\ Westphal) The extension of
		the Hubble diagram. II. New redshifts and photometry of
		very distant galaxy clusters: First indication of a
		deviation of the Hubble diagram from a straight line.
		{\it Astrophys.\ J.} {\bf 221,} 383--394.
  \item[1982] (With G.\ A.\ Tammann) Steps toward the Hubble constant.
		VIII. The global value.
		{\it Astrophys.\ J.} {\bf 256,} 339--345
  \item[1984] (With G.\ A.\ Tammann) The Hubble constant as derived from
		21 cm linewidths.
		{\it Nature} {\bf 307,} 326--329. 
  \item[1985] (With B.\ Binggeli \& G.\ A.\ Tammann) Studies of the Virgo
		cluster. V. Luminosity functions of Virgo cluster galaxies.
		{\it Astronom.\ J.} {\bf 90,} 1759--1771.
  \item[1990] (With J.-M.\ Perelmuter) The surface brightness test for
		the expansion of the universe. I. Properties of Petrosian
		metric diameters.
		{\it Astrophys.\ J.} {\bf 350,} 481--491.
  \item[1999] The first 50 years at Palomar: 1949--1999. The early years
		of stellar evolution, cosmology, and high-energy
		astrophysics.
		{\it Annu.\ Rev.\ Astron.\ \& Astrophys.} {\bf 37,}
		445--486.
  \item[\yr] (With A.\ Saha, G.\ A.\ Tammann, L.\ Labhardt, F.\ D.\ 
		Macchetto \& N.\ Panagia) Cepheid calibration of the peak
		brightness of type Ia supernovae. IX. SN 1989B in NGC 3627.
		{\it Astrophys.\ J.} {\bf 522,} 802--838.
  \item[2001] (With L.\ M.\ Lubin) The Tolman surface brightness test for
		the reality of the expansion. IV. A measurement of the
		Tolman signal and the luminosity evolution of early-type
		galaxies.
		{\it Astronom.\ J.} {\bf 122,} 1084--1103. 
  \item[2006] (With G.\ A.\ Tammann, A.\ Saha, B.\ Reindl, F.\ D.\
		Macchetto, \& N.\ Panagia) The Hubble constant: A summary
		of the Hubble Space Telescope program for the luminosity
		calibration of type Ia supernovae by means of Cepheids.
		{\it Astrophys.\ J.} {\bf 653,} 843--860.
  \item[2008] (With G.\ A.\ Tammann \& B.\ Reindl) The expansion field:
		The value of $H_0$.
		{\it Astron.\ \& Astrophys.\ Rev.} {\bf15,} 289--331.
  \item[2010] The Tolman surface brightness test for the reality of the
		expansion. V. Provenance of the test and a new
		representation of the data for three remote Hubble Space
		Telescope galaxy clusters.
		{\it Astron.\ J.} {\bf 139,} 728--742.
  \item[\yr] (With B.\ Reindl \& G.\ A.\ Tammann) The linearity of the
		cosmic expansion field from 300 to 30,000 km/s and the
		bulk motion of the local supercluster with respect to
		the cosmic microwave background.
		{\it Astrophys.\ J.} {\bf 714,} 1441--1459.

  \item[{\rm Books}]
  \item[1961] The Hubble Atlas of Galaxies.  Carnegie Institution of
		Washington.
  \item[1975] (With M.\ Sandage \& J.\ Kristian, eds.) Stars and Stellar
		Systems. IX. Galaxies and the Universe.  The University
		of Chicago Press.
  \item[1981] (With G.\ A.\ Tammann) A Revised Shapley-Ames Catalog of
		Bright Galaxies.  Carnegie Institution of Washington.
  \item[1988] (With J.\ Bedke) Atlas of Galaxies Useful for Measuring
		the Cosmological Distance Scale.  NASA SP--496.
  \item[1994] (With J.\ Bedke) The Carnegie Atlas of Galaxies.  Carnegie
		Institution of Washington.
  \item[2004] Centennial History of the Carnegie Institution of
		Washington, Vol.\ 1: The Mount Wilson Observatory.
		Cambridge University Press.
\end{description}

\noindent
Sandage was an Editor of the {\it Annual Review of Astronomy and
Astrophysics} from 1990 Volume 28 to 2004 Volume 42.

\end{document}